# Interactive Museum Exhibits with Microcontrollers: A Use-Case Scenario

Lok Wong
Univ. of Calif, San Diego
La Jolla, CA, USA
lywong@ucsd.edu

Shinji Shimojo
JGN-X, National Institute of Communication and Technology
Osaka, Japan
sshimojo@gmail.com

Yuuichi Teranishi, Tomoki Yoshihisa
Cybermedia Center, Osaka Univ., Mihogaoka, Japan
teranisi@cmc.osaka-u.ac.jp,
yoshihisa@cmc.osaka-u.ac.jp

Jason H. Haga
Cyber-physical Cloud Research Group, AIST
Tsukuba, Japan
jh.haga@aist.go.jp

## ABSTRACT
The feasibility of using microcontrollers in real life applications is becoming more widespread. These applications have grown from do-it-yourself (DIY) projects of computer enthusiasts or robotics projects to larger scale efforts and deployments. This project developed and deployed a prototype application that allows the public to interact with features of a model and view videos from a first-person perspective on the train. Through testing the microcontrollers and their usage in a public setting, it was demonstrated that interactive features could be implemented in model train exhibits, which are featured in traditional museum environments that lack technical infrastructure. Specifically, the Arduino and Raspberry Pi provide the necessary linkages between the Internet and hardware, allowing for a greater interactive experience for museum visitors. These results provide an important use-case scenario that cultural heritage institutions can utilize when implementing microcontrollers on a larger scale, for the purpose of increasing visitors' experience through greater interaction and engagement.

## Categories and Subject Descriptors
C.3 Special-purpose and application-based systems: Microprocessor/microcomputer applications. J.5 Arts and humanities.

## 1. INTRODUCTION
The Internet of Things (IoT) refers to a global infrastructure that is interconnecting both physical and virtual objects (e.g. sensors, devices, actuators, etc.) based on evolving technologies and allowing them to transmit and receive data [1, 2]. This connectivity allows an entirely new level of information access, retrieval and interaction possible as the physical world becomes connected to the Internet. Business enterprises and industry sectors are moving ahead with deploying this technology in a wide variety of ways [3] and there is a sense that these efforts are moving rapidly away from other sectors of society [4].

Yet, it is important to note that IoT will touch all sectors of society as it becomes more ubiquitous. Cultural heritage institutions have much to gain by implementing IoT technologies and create applications that enhance a visitors' experience. As the mode of interactions become more integrated with digital/virtual sources, the benefit of connecting to visitors via technology is impactful in allowing these institutions to retain visitor attention and interest, as well as reaching a larger audience [5]. Although artists use microcontrollers to engage visitors, these efforts typically only support or convey a single artistic vision [6]. Museum institutions as a whole only recently have implemented IoT technologies for public education and engagement on a broader level [7]. The National Museum Wales became the first in the world to embark on a full-fledged IoT pilot in 2014 [8], suggesting that museums that are crafting digital strategies and determining how to use IoT would benefit from lessons learned in use-cases, furthering the development and deployment of these technologies in the arts and humanities domains. With this motivation in mind, this project provides an important use-case scenario that can be used as a guide by cultural heritage institutions to create more engaging and interactive exhibits using technology.

## 2. BACKGROUND AND OBJECTIVE
In recent years there has been a shift in the concept of what a museum visit is. One of the main factors is the use of technology. The United States based National Endowment for the Arts conducted a recent survey that found 75% of Americans used electronic devices to experience art, while just 33.4% attended one of seven art events the same year [9]. With this trend there is a growing consensus among museum professionals that integrating new technologies with cultural heritage will become more important in engaging and educating museum visitors [10].

The San Diego Model Railroad Museum released a miniature train model that reflects the state and beauty of Balboa Park when it was first founded as part of its centennial celebration [11]. The museum has found it challenging to reach a particular demographic - teenagers. The museum is well loved by train fanatics and parents who bring their young children to enjoy it. However, the teenage group is largely missing from visiting and participating in the museum and its events. In an effort to increase teenagers' engagement, the museum wanted to incorporate modern technology with the miniature train model. Leveraging on the concepts of IoT, the museum hoped to use microcontrollers to increase the interaction between the model and the users, allowing the train model to become more accessible and enjoyable for people of all ages, especially the teenage demographic.

Two objectives were conceived at the start of the project: 1) to allow users to connect to the miniature model through the Internet and 2) to allow people to see videos of the models. The first idea gives visitors control over some physical aspects of the model, making it more engaging and interactive. The second idea would allow visitors to stream live, real-time video on the Internet of the first-person view from the train, giving visitors and potential visitors a unique perspective. By integrating different types of microcontrollers into the model, these ideas were used to create and test a working prototype in a public area that would simulate the situation at the museum.

## 3. PROJECT DETAILS AND TESTING
## 3.1 Controlling devices through the Internet

As part of the interactive experience, visitors would be able to interact with the fixtures and appliances on the train model directly using the Internet. Visitors were given control over the sounds and lights of the model, which was consistent with some of the concepts of IoT involving connecting physical objects to the Internet.

Initially, existing commercial, home-automation products were considered as the hardware component that connects the fixture to the software interface (e.g. Belkin WeMo, Smarthome, etc). However, this approach suffered from security concerns such as weak authentication protocols and providing attackers with a means to relay commands to other devices. It was also cost prohibitive, especially for small-scale museums that have operating budgets less than $5 million [12].

As a solution, microcontrollers were found as a suitable alternative to these commercial products in connecting appliances to the Internet, serving as the bridge between physical fixtures and existing networks. For this task, the Arduino Uno was selected because of its ability to manipulate devices and its well-developed open-source library [13].

### 3.1.1 Technology

To allow users to control the appliances in the train model several links must be made. There must be links between the appliance and Arduino, between the Arduino and server, and between the server and browser. The appliance in this case is the light or sound device and the browser is the web interface. Figure 1 shows the general architecture and overall set-up of these different components.

The PowerSwitch Tail II is a piece of hardware that receives a signal from the Arduino to control the voltage of appliances, thus regulating the link between the appliance and Arduino [14]. This allows the Arduino to turn on and off appliances; specifically in this case it was used to control a light fixture in the train model. To bridge between the server and Arduino, Transmission Control Protocol (TCP), was used to push data from a Node.js server to the Arduino. Node.js was used as the server platform because of its support for fast and real-time communication between different servers, which is essential for a fast and scalable network for the users [15].

To create the link between Node.js server and web browser, the JavaScript library Socket.IO was used to push data onto the server. Socket.IO enables real-time communication between the browser and the server that is essential in creating real-time feedback that the users could see [16]. To facilitate user interactions, a simple web browser interface was implemented such that users can send on/off signals to the server.

Another interactive component of the train model was to control sound fixtures. Essentially, music from the era will be played in the background, enriching users' experiences and their knowledge of the time period. The Arduino itself does not have an audio outlet or SD-card capability, but it can be deployed as a MP3 music player and storage device through an add-on module called the Music Shield [13]. The Arduino could then play music through speakers, with low power requirements.

### 3.1.2 Testing

A prototype, simple user-interface coupled to a basic Node.js

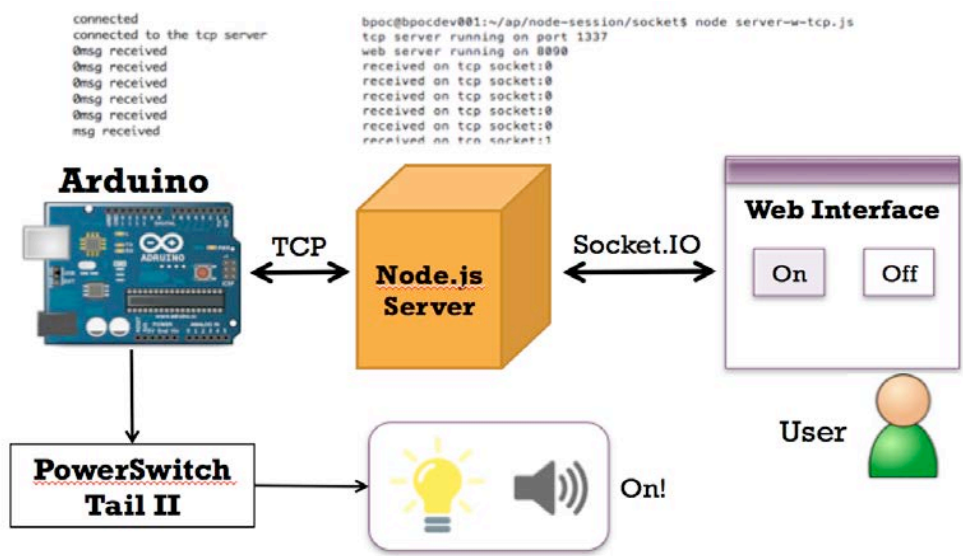

**Figure 1. Diagram outlining the general architecture of the hardware and software for the Arduino microcontroller setup.**

server that simulates the train model set-up was developed for testing. A light fixture was connected to the PowerTail Switch II. In the prototype, users click the button on the user-interface, as shown in Figure 2, hosted by the server five times to turn the light on. Clicking the button one additional time turns the light off. A similar implementation was created for the audio. This overly simplified interface facilitated the testing of the prototype, but we expect a more comprehensive design that includes all controllable physical features of the model to be presented in unified interface in the future.

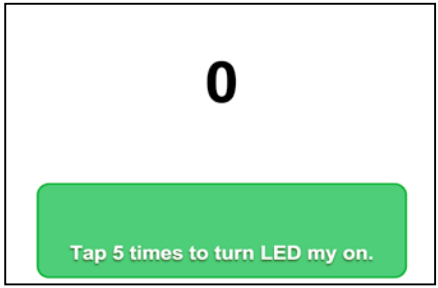

**Figure 2. Diagram of the prototype user interface. The number displayed corresponds to the number of button taps the user inputs.**

When tested, the prototype works as expected for small number of users (five to six people). However, when the group of users increases to more than seven users, the website and servers could not maintain a uniform count, and could not send the correct signal to the Arduino, despite the scalable aspects of Node.js. We are investigating this problem and will address this in the future.

## 3.2 Real-time Video from Miniature Trains

To stream real-time video from a miniature train, allowing visitors to view the local model scenery from the perspective of the train, depended on the form factor of the train itself. There are physical limitations imposed upon the technology that could be used because of the train size and dimensions. It was not feasible to put a high quality camera or full-size computer onto the train. In addition, the train size, as well as the physical layout of the miniature train model prevented the installation of a hardwired constant power supply.

To overcome these limitations, the technology that was chosen must be agile, support the camera hardware, and be power-efficient. In searching through the currently available technologies, the health care field has assessed microcontrollers as potential medical electronics as it is cost-efficient and power-efficient [17]. Because of this, microcontrollers were chosen for this project as the most suitable solution to all the constraints.

### 3.2.1 Technology

As one of the most prevalent microcontrollers, the Raspberry Pi has the capacity to act as an inexpensive server, while supporting an extensive amount of add-ons, such as a camera [18]. By adding a camera module onto the Raspberry Pi board, the microcontroller could host a server, relaying footages captured by the module onto the Internet and enable users to view it.

Due to the compactness of the Raspberry Pi, it is agile enough to fit onto a miniature train. As it can also be powered with external batteries via USB, the Raspberry Pi circumvents the power limitation of the miniature model. Figure 3 displays the configuration of the Raspberry Pi on the miniature train. The camera is mounted on the train roof in the front to capture a view of the model landscape as the train moves. The Raspberry Pi was placed behind the camera in the interior of the miniature train.

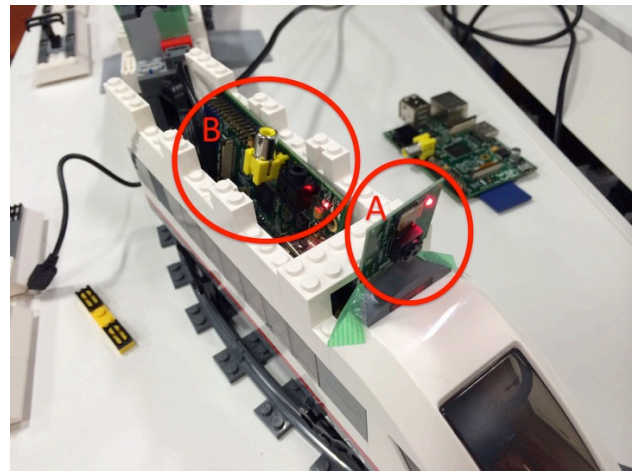

**Figure 3. Prototype of the camera set-up on a train made from Legos. Note that the camera appears at the front in the first train car (A), followed by the Raspberry Pi (B), and the battery is in the second train car (not pictured).**

External batteries were placed further behind in the adjacent train car compartment.

The Raspberry Pi, in essence, acts as a server to stream video and also as the camera in which it captures the video using the Raspberry Pi camera module and records the video using MJPG-streamer, a well-establish method to stream video [19].

To provide personalized views for respective visitors, the video streams captured by the Raspberry Pi cameras were delivered to user devices (tablets, smartphones, etc.) along with a public monitor device. However, the Raspberry Pi does not have enough computational/network power to serve multiple streams to many devices. Furthermore, user devices require different quality of video streams. For example, a PC or a public monitor device will display videos with 30 fps (frames per second) while smartphones can only display videos with 15 fps. If only one type of video stream quality is used, redundant data transmission would occur. In the above example, if there exists only 30 fps video stream, the smartphone cannot display half of the received video stream data. The wireless network resources would be needlessly consumed by the excess data transmission. If a 15 fps video stream is used to address this problem, then the PC/public monitor device will suffer from poor video streaming quality.

Therefore, we implemented an adaptive video stream delivery system, which adjusts the computational/network resources to serve and control the delivery quality for the respective devices. As shown in Figure 4, this system has source servers that deliver the video streams captured from the Raspberry Pi camera and destination servers that serve the video streams to each user device with required qualities adaptively. The quality of the video is decided according to the device's specifications or the delivery status. The stream data delivery follows an algorithm called LLF [20] that can exchange different qualities (transmission cycle) of video streams effectively in a peer-to-peer manner. Each user's device can receive a MJPEG video stream, which is supported by most of the web browser implementations including ones on smartphones.

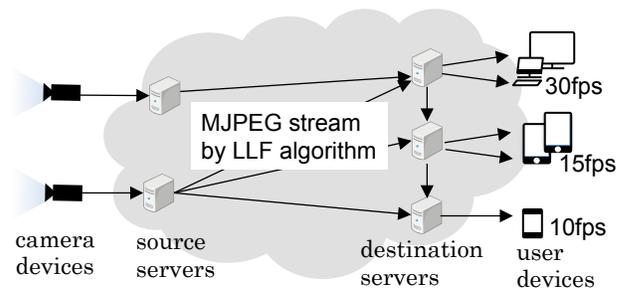

**Figure 4. Diagram illustrating the adaptive video stream delivery system. Source and destination servers are able to stream different quality video depending on the frame rate (frames/sec) required by the user device.**

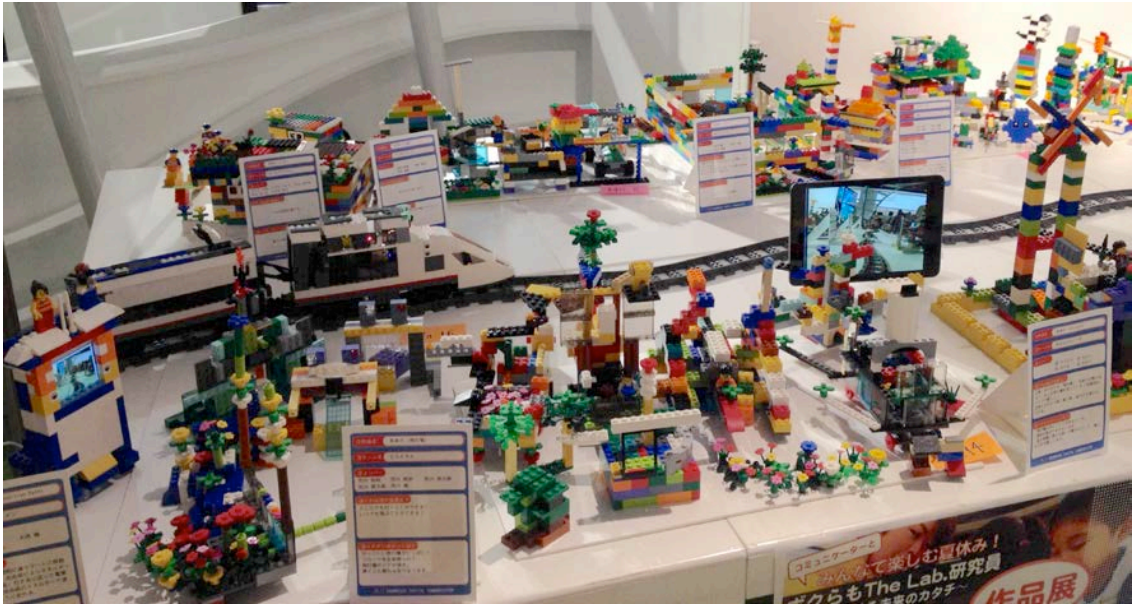

**Figure 5. Overall view of the public exhibit of the Lego models. The train with Raspberry Pi camera moves through the center of the models. Both the 10" tablet (right) and a smartphone mounted in a structure (far left) are visible, with streaming video from the train's position.**

### 3.2.2 Testing

In collaboration with Kobe Institute of Computing, the technology was tested through a publicly held exhibit at The Lab in the Grant Front Osaka (Kita Umeda, Japan). The technology was tested on and used to enhance the experience of one of The Lab's public events. As part of the event, children were asked to create Lego models that display their visions of future cities. Our prototype was used to show the children their models through the lens of the train. Several different display devices were placed throughout the models, including a 10" tablet, a mini 7" tablet, and a smartphone. A larger display monitor was also placed behind the models to give visitors further away from the model the opportunity to view the streamed video. The current prototype did not stream video to a visitor's personal device, thus, there was limited usability by the visitors. Figure 5 and 6 show the technology on display during the public event with the video streaming displays and functional train-mounted camera.

The adaptive video stream delivery system was implemented on the high-speed experimental testbed network called JGN-X [21]. While the server resources on JGN-X were used to distribute the video streams, the devices in the model exhibit at The Lab were connected via locally available, public Wi-Fi to receive the video streams. When the server on the Raspberry Pi is run on public Wi-Fi, the quality of the video streams to the devices was not optimum and the resulting blurry image was difficult to view. By optimizing the Wi-Fi settings and strengthening the signal, the quality of video improved, however, this was only a partial solution and the image quality could be improved further. In spite of this, generally the adaptive streaming functioned properly and streamed different quality videos to each of the devices present.

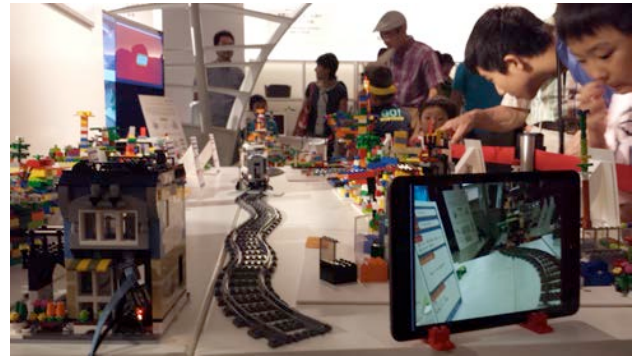

**Figure 6. Public exhibit demonstrating the Lego models with the train with Raspberry Pi camera. Note on the right is a monitor displaying the live video stream from the train.**

## 4. DISCUSSION

This use-case has successfully created working prototypes that accomplished the objectives of the project. While further development will be needed to address some issues, the technology implemented demonstrates the feasibility of this approach, especially for cultural heritage institutions as a whole. As mentioned in the beginning of this paper, this effort was carried out considering that these technologies will be deployed in a museum that traditionally lacks technical infrastructure and included multiple stakeholders (i.e. administrators, educators, curators, developers, etc.). This work is an important "lessons learned" for any cultural heritage institution that is considering a similar deployment.

While the microcontrollers work well for small-scaled projects, their ability to scale and operate dependably in a large-scale environment is still under investigation. Institutions should have a digital strategy that factors in large number of users and the deployment should be capable of adjusting to it. With this project, the scalability issue will be first addressed by deploying a dedicated server for Node.js that will be able to handle a large number of user requests. The simplicity of the user-interface makes further optimization of the web-browser for scalability impractical. However, future implementations will need to maintain this simplicity for maximum usability by the public, hence satisfying the mission of cultural heritage institutions to broaden visitor engagement and education.

One additional consideration for the user interface, is the issue of dominance or user priority when allowing the public to interact with an exhibit. When the public controls a fixture in the train model, it is necessary to have an approach to prioritize which user has control of the physical object at a given time. There are several different approaches; including first-come first-served with a time delay before the next user has access. Another is having a minimum number of button taps before the fixture activates, which would encourage multi-user cooperation. An additional option is to implement a crowd sourced voting system that uses the most popular visitor selection from a fixed time period. These approaches and others will be investigated in future work, but are important to discuss with the stakeholders during the design and deployment of these technologies.

A working prototype using the Raspberry Pi as a server and device to stream video from a moving train was also implemented. This prototype was tested at a public event and exhibition to assess the feasibility of using it in a public setting. Although successful, issues were found with regard to microcontrollers and the streaming technology involved. Some concerns have arisen with the power capacity of the microcontroller as it relies on external batteries. Currently, the batteries have an operating lifetime of about 3 hours, which is suitable for short-term demonstrations. Longer demonstrations will require extending the operating lifetime. One possible solution is inductive or wireless charging [22], but this is currently a developing technology and would add substantial cost to the deployment that would be beyond most cultural heritage institutions' budgets.

As mentioned in a previous section, public Wi-Fi was not sufficient for good quality video streaming, most likely due to the high number of users on the network. Public Wi-Fi was selected for this project as it provided easy access to the Internet without the need for procurement and installation of additional hardware. The miniature train model would benefit from a dedicated, private Wi-Fi network or Ethernet-based Internet access, but this may be cost prohibitive to install. As a lesson learned, stakeholders involved in similar deployments should carefully consider both the time that the exhibition will be available to the public (temporary or permanent) and the type of data that the public will consume (text, audio, video, etc.) and the availability of funds. This again alludes to the fact that cultural heritage institutions should have a comprehensive digital strategy that is flexible enough to accommodate different scenarios.

The successful development and deployment of this prototype in a public space demonstrates the great potential in improving the visitor experience by cultural institutions. The large amount of positive feedback the prototypes have generated at the public exhibit space suggests that these products could be used at other public institutions, with this use-case as a guide. This project shows the applicability of IoT in cultural heritage institutions and additional use-cases such as this one will continue to advance the development of new applications of IoT in creating enhanced interactive experiences that engage and educate the public at large.

## 5. ACKNOWLEDGMENTS

The authors would like to acknowledge support from the UCSD Pacific Rim Experiences for Undergraduates program (PRIME NSF INT 0407508 and NSF OISE 0710726), PRAGMA, NICT, and the San Diego Model Railroad Museum.

The authors would also like to acknowledge support from Balboa Park Online Collaborative, from the professors and students of Osaka University and Kobe Institute of Computing for their work in preparing and during the exhibit, and from the staff of The Lab in Grand Front Osaka for hosting the exhibition.